# Can LHC observe an anomaly in t$\bar{t}$Z production?

François Richard

Laboratoire de l'Accélérateur Linéaire, IN2P2–CNRS et Université de Paris–Sud,
Bât. 200, BP 34, F–91898 Orsay Cedex, France.

**Abstract**
The cross section for t$\bar{t}$Z production at the 7 TeV LHC has been measured. For the first time it therefore becomes possible to measure Z couplings to top quarks. Interpreting the notorious LEP1 anomaly on Z couplings to b quarks in terms of an extra-dimension model, one expects an excess, as large as a factor 2, on the t$\bar{t}$Z rate which may already become significant with the LHC at 8 TeV data. Other schemes are reviewed which also predict deviations of t$\bar{t}$Z couplings. The size of the effect and a dramatic increase of the top quarks with the right chirality should confirm our interpretation of the underlying mechanism. Complementary signals observable at LHC are also briefly reviewed.

# 1. Introduction

In a previous paper [1] the deviation observed at LEP1 on AFBb has been interpreted within the Randall Sundrum (RS) model with an extra dimension. To achieve this, it was assumed that the b quark with right chirality bR couples strongly to the Kaluza-Klein (KK) vector states within RS. As pointed out in the same paper one also predicts radical changes for top physics at ILC within the same scheme. This feature is quite intuitive given that this model generically predicts a much stronger coupling of the KK states to tR.

What was not clearly foreseen however is that LHC measurements, even with the modest luminosity collected so far, could become sensitive to such effects. A recent paper [2] from CMS indicates that this is not so and that LHC offers good prospects on this measurement provided that the deviation predicted by the RS model is large enough. In this note it will be shown that an enhancement as large as 2 can be foreseen. From the expected cross section reported for 7 TeV data in [2] and the clean signal observed with 3 lepton final states, it becomes conceivable to observe a significant signal with the data collected so far.

# 2. The model

In [1], it was assumed that the effect observed at LEP1 on the b forward-backward asymmetry AFBb was due to the mixing between Z, its recurrence Zkk and a Z', the latter being an extra boson predicted within extended symmetries which are needed to accommodate the LEP/SLD/TeVatron precision measurements and RS contributions. The dominant contribution was due to the latter since one has the freedom to assume that its coupling is much larger than the Z coupling. Given the gauge group used, the enhanced coupling only affects the right chirality component bR and this is precisely what is needed to explain the observation of LEP1.

At the time of our publication, there was no significant deviation on Rb related to the partial width of Z into b quarks. This was explained by an accidental cancellation between the contribution due to bR and a smaller contribution due to bL. Since then, it was understood that the theoretical prediction for Rb receives extra theoretical corrections [3] meaning that both Rb and AFB deviate at the ~2.5σ level from the SM. This feature is easily accommodated within our scheme by reducing the variations of the couplings gRb and gLb but does it not alter the main conclusion.

Quantitatively, before one had dgRb/gRb=-29% and dgLb/gLR=-1.9% and now one has dgRb/gRb=-21% and dgLb/gLb=-0.24% [4]. In our model one can write:

$$\frac{dg_{Rb}}{g_{Rb}} = \left(\frac{M_Z}{0.4 M_{KK}}\right)^2 \left[1 + \frac{1}{4F(c_{bR})} + \frac{g_Z^{2\prime} I_{3R}^{bR}}{g_Z^2 Q_b s_W^2}\right] F(c_{bR})$$

and:

$$\frac{dg_{Lb}}{g_{Lb}} = \left(\frac{M_Z}{0.4 M_{KK}}\right)^2 \left[1 + \frac{1}{4F(c_{bL})} - 0.9\right] F(c_{bL})$$

where F(c) is a function describing the distribution of a given flavor in between the two branes of the RS model. Having fixed Mkk=3 TeV and gz'/gz~5 (this choice is close to the unitary limit g²z'~4π which corresponds to a ratio of 5.5) one had a solution with cbR=0.49 and cbL=0.35. The new SM solution can be trivially adjusted with cbR=0.50 and cbL=0.47.

Recall that gLb=(-0.5-Qbs²w)/swcw=-1.00 and gRb=-Qbs²w/swcw=0.18 with Qb=-1/3 and s²w=0.231.

Using the formalism of our paper, one can write:

$$\frac{dg_{Rt}}{g_{Rt}} = \left(\frac{M_Z}{0.4 M_{KK}}\right)^2 \left[1 + \frac{1}{4F(c_{tR})} + \frac{g_Z^{2'} I_{3R}^{tR}}{g_Z^2 Q_t s_W^2}\right] F(c_{tR})$$

The function F(ctR) should be, in absolute value, very large to explain the large Yukawa coupling of the top quark. This large value implies that the variation of gRt is much larger than for gRb.
Recall that gLt=(0.5-Qts²w)/swcw=0.82 and gRt=-Qts²w/swcw=-0.37 with Qt=2/3 and s²w=0.231.
To understand the correlation between $t\bar{t}Z$ and the LEP1 measurement, one can simply write that:

$$\frac{dg_{Rt}}{g_{Rt}} \Big/ \frac{dg_{Rb}}{g_{Rb}} \sim \frac{Q_b I_{3R}^{tR}}{Q_t I_{3R}^{bR}} \frac{F(c_{tR})}{F(c_{bR})}$$

As already mentioned, LEP1 gives F(cbR)~-0.25. We have also seen that when the gRb measurement varies, F(cbR) needs to be adjusted but the ratio (dgRb/gRb)/F(cbR) remains the same. One therefore concludes that dgRt/gRt only depends of F(ctR). This function varies typically between -5 and -10 for ctR chosen between 0 and -1. To avoid having light KK quarks which are already excluded by LHC, one should restrict to ctR>-0.5.
gLt has an almost negligible variation given that, from weak isospin symmetry, one predicts ctL=cbL. Quantitatively one has dgRt/gRt=-3.5 (choosing ctR=-0.5) and dgLt/gLt=-0.24%. At LHC, the enhancement of the $t\bar{t}Z$ cross section with respect to the SM is trivial to compute since, for top pairs, QCD predicts an equal amount of tL and tR, hence a $t\bar{t}Z$ cross section proportional to gRt²+gLt².
Figure 1 shows the dependence of the predicted RS enhancement in terms of -ctR together with the result from CMS (centered at ctR=-0.5) which is commented below. As commented in the next section, one should consider the two possible signs for the ratio I3Rt/I3Rb and figure 1 shows the rather strong change between these two solutions.

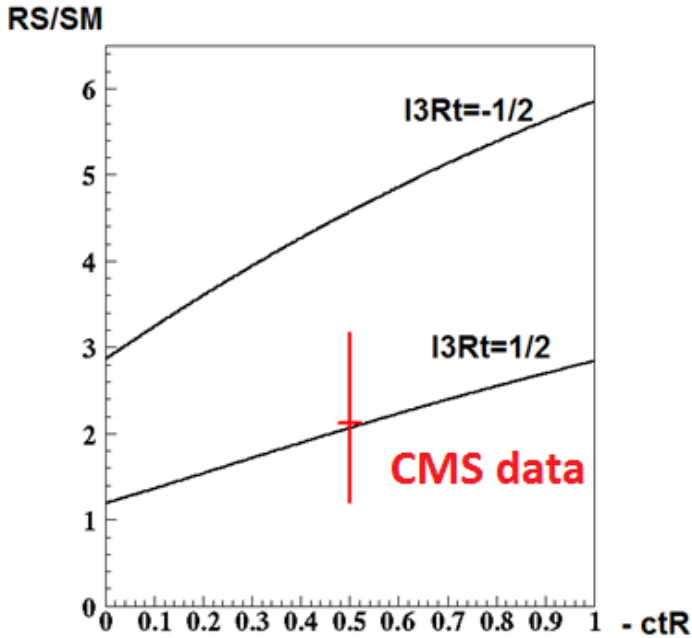

Figure 1: The lower curve gives the enhancement of the LHC $t\bar{t}Z$ rate due to the RS contribution vs the value of -ctR, where ctR enters in the function F(ctR) defined in the text. The red bar gives the measured $\pm\sigma$ interval for the ratio between the CMS measurement and the prediction from NLO. The upper curve corresponds to an opposite choice for the tR isospin third component.

# 3. Interpretations

It is fair to say that there is some flexibility in above prediction and that, e.g. one can compensate a variation of the Z' coupling by a change of the KK mass without consequences for LEP1 observables. This flexibility precludes a precise prediction of Mkk. Note however that a lower value of Mkk is usually discarded on the basis of electroweak precision measurements (S parameter). Mkk cannot be increased since our choice for the ratio gz'/gz is already close to the unitary limit. This therefore allows to speculate that the first KK resonance might be within future reach of LHC. The RS production mechanism for top pairs is discussed in [5] where one assumes that the AFBt anomaly observed at Tevatron calls for gluon KK excitation with mass ~2-3 TeV. This prediction is not yet eliminated by LHC searches given the large width predicted for such a resonance and also given the experimental limitations for reconstructing energetic tops, the so-called 'boosted top' problem.

Another promising channel discussed in [6] would be the observation of a Z'/Zkk decaying into WW or a Wkk decaying into ZW. So far the WW channel is cleanly identified in the leptonic channel which does not allow precise mass reconstruction. The ZW channel, with Z into lepton pairs, gives a precise mass reconstruction.

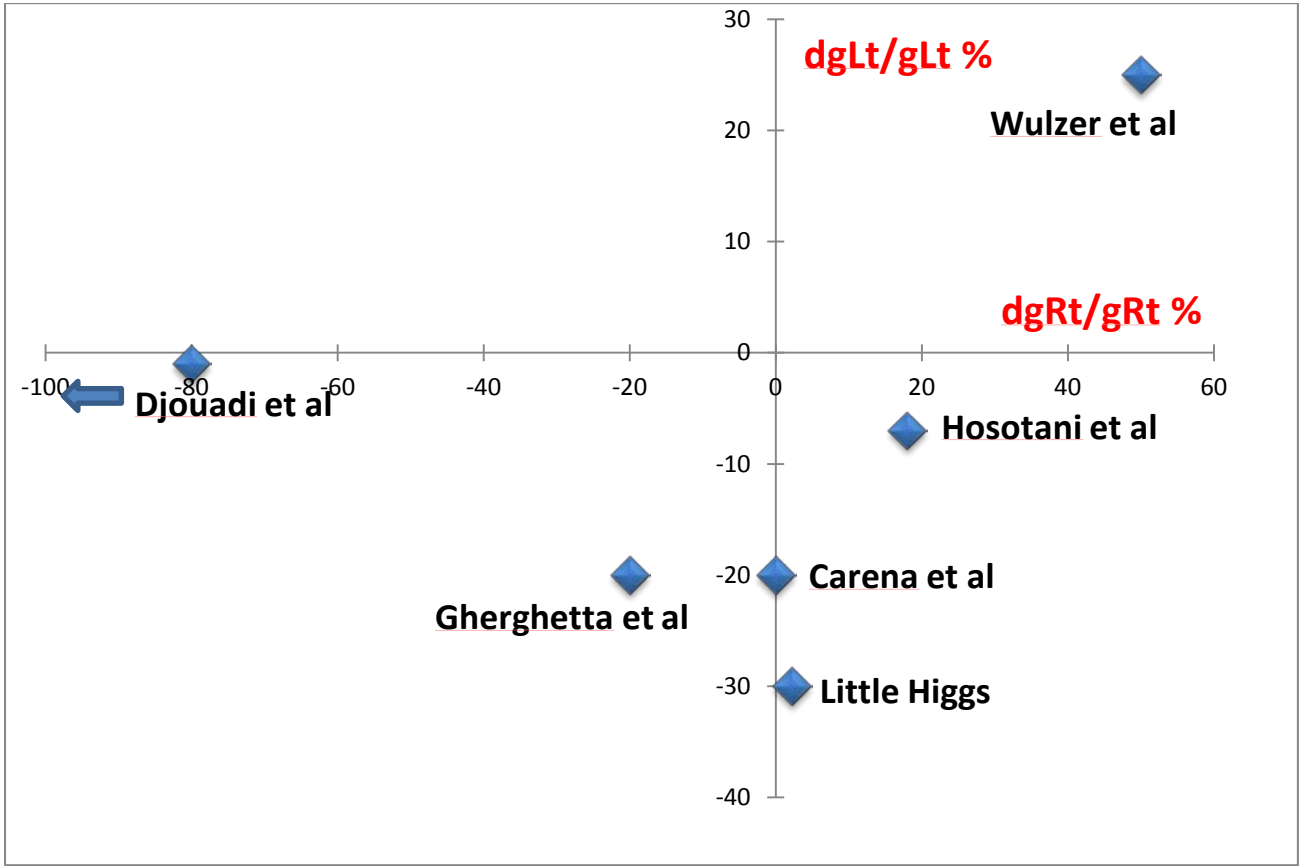

Figure 2 : Variation of tL and tR couplings to Z within models referenced in the text. In this paper our solution can be far off in the negative direction of the gRt axis as the arrow indicates next to the 'Djouadi et al' solution (which corresponds to ctR=0.35).

In our prediction one takes $I_{3R}^{tR}$ =1/2 and $I_{3R}^{bR}$ =1/2 but one cannot exclude other values. In [1] one had explored the solution with $I_{3R}^{bR}$ =-1/2 which requires dgRb/gRb=2.2. Such a solution works taking cbR=0.35 and is still acceptable from the LEP1 data.

$I_{3R}^{tR}$ =0 would of course eliminate any large enhancement while $I_{3R}^{tR}$ =-1/2 would increase it further (as shown in figure 1) since dgRt/gRt would become positive. If a significant enhancement is observed, one has $I_{3R}^{tR}=\pm 1/2$ and therefore there should be quark partners of tR and bR called b'R and t'R. This type of quark is a KK particle with no 'zero mode' meaning that it does not receive its mass from the Higgs mechanism. It can accordingly be very heavy and therefore present searches performed at LHC which reach mass limits of 500-700 GeV do not rule out this possibility. With LHC running at 13 TeV, one may hope to observe a heavy quark which would confirm the underlying mechanism.

One can of course propose other extra-dimensional models [7] which consider both Z-Z' mixing and top mixing with new heavy quarks. None of them provide a large increase in the $t\bar{t}Z$ rate. They can modify the couplings of Z to tL and tR as shown in figure 2.

The same occurs in schemes invoking b mixing with vector-like quarks alone[8]. It is also possible to reproduce the AFBb and Rb measurements but there is no impact on the top sector since a large mixing of tR with a heavy quark would impact on b->sγ.

The Little Higgs (LH) scheme[9] will tend to give a reduction in rate which can be important. For instance assuming t-T quark mixing and with mT=1 TeV and $\lambda_T$=2 one generates dgLt/gLt~-20%. Gauge boson mixing can generate a similar value. gRt receives no contribution from T-t mixing and an almost negligible value from gauge boson mixing. Hence LH predicts that the $t\bar{t}Z$ rate could go down by up to a factor 2 with respect to the SM. This prediction is therefore very much contrasted with our model.

A final example[10] is a composite approach where the composite Higgs is realized as a Goldstone boson associated with a symmetry breaking, the top quark can also have partners which mix with the top quarks modifying the couplings to tL or tR. The solution from [10] can increase the $t\bar{t}Z$ rate by ~1.7, noting that, contrary to our model, the dominant component remains tL.

This brief and incomplete survey is illustrated by figure 2.

One concludes that a large excess (~2) in the $t\bar{t}Z$ rate, easily observable at LHC, is favored within our scheme deduced from AFBb at LEP1 which predicts a very large variation of gRt. It is also clear that, in all instances, measuring precisely and separately the Z coupling to tL and tR will allow a good discrimination within the jungle of proposed models. It has recently been shown [11] that this is indeed possible at ILC with relative errors at the % level.

## 4.The data

CMS has published[2] a first result using the data collected at 7 TeV. They use two samples, one with 3 leptons compatible with $t\bar{t}Z$ and the other with two like sign leptons compatible with $t\bar{t}W$ (or $t\bar{t}Z$ with one lepton missed). Since the later is not directly related to $t\bar{t}Z$, it will be ignored.

The observed cross section is $\sigma_{t\bar{t}Z}=0.28^{+0.14}_{-0.11}(stat.)^{+0.06}_{-0.03}(syst.)$ pb while for the NLO prediction one has $\sigma_{t\bar{t}Z}=0.137^{+0.012}_{-0.016}$ pb. This value corresponds to ~0.1% of the top SM cross section. Given the number of events predicted in [2] the reconstruction efficiency corresponds to ~0.5% which includes the Z and top leptonic branching ratios. The expected number of 3 lepton events is ~2.7. Clearly this result is statistically compatible both with the SM and the RS prediction.

Using the 7 TeV data, ATLAS has published [12] a 95% credibility upper limit, 0.7pb, on the $t\bar{t}Z$ channel using 3 lepton final states. One candidate was observed while the expected number is 0.85±0.04(stat)±0.14(syst) with a background of 0.28 ±0.05(stat)±0.14(syst). Again this result does not allow to decide between SM and RS.

## 5.Additional features and prospects

The relative amount of tL and tR is equal in the QCD SM top pair production process. For $t\bar{t}Z$ there is a marked asymmetry since Z couples preferentially to tL (gLt/gRt~2.2). In the SM one therefore expects that 83% percent of $t\bar{t}Z$ events contain tL. Noteworthy is the fact that, for tL, the lepton spectrum is markedly softer than for tR. The reason is simple: helicity conservation imposes that for tL the W is preferentially emitted backward in the top rest system and therefore, after the Lorentz boost effect, one expects to observe soft leptons in the laboratory. The opposite is true for tR. Detection should therefore be more favorable for tR.

Within our version of RS however tR regains the majority, with ~60% (assuming ctR=-0.5) of the tops being tR, and therefore the lepton from top decays should be on average more energetic, a feature which should become observable with more data. *Remarkably this prediction is unique to our model and this feature should be absent within the various other models summarized by figure 2.*

One can already notice that this increase in the tR component may result in an increased detection efficiency which can amplify the excess with respect to the SM prediction. In alternate schemes discussed previously there is the possibility that the tL components is modified, in certain cases significantly increased, and since this is the dominant part at the SM one will not observe a significant alteration of the tL dominance.

With 5 times more data and an enhanced (~1.5) cross section at 8 TeV, CMS should be able observe a ~2 to 3$\sigma$ excess on $t\bar{t}Z$ as predicted by our model with ctR=-0.5. Note that this estimate assumes that the systematical errors will not decrease with additional data which is somewhat pessimistic. If this effect is indeed observed, it will be important to look at the distribution of the lepton coming from top decays to establish that it is not mainly coming from tL decays.

Note that, within the RS model, no measurable effect is predicted for the channel $t\bar{t}\gamma$ in the absence of a Z-Z' mixing type mechanism.

In the future, with about 50fb-1 luminosity cumulated at 13 TeV and assuming a factor 2 increase in the $t\bar{t}Z$ cross-section, the statistical error would reach about 5% which allows to demonstrate the excess beyond any doubt. However if the effect turns out to be smaller, a major limitation to reach the 5$\sigma$ evidence requires to control the systematical errors to better than 10% which may not be trivial.

## 6. Conclusion

In this note it was shown that, from the deviation observed at LEP1, one is entitled to expect a factor ~2 enhancement in the rate of $t\bar{t}Z$ observed at LHC. This strong enhancement is provided by Z-Z' mixing where Z' is a KK boson appearing within the RS phenomenology with an expected mass ~3 TeV. Given the excellent quality of the LHC detectors and the clean signal provided by $t\bar{t}Z$ final state with Z decaying into two leptons, such an enhancement could be already observed with present data.

In the future, the statistical error on the $t\bar{t}Z$ cross section will become negligible with respect to the presently quoted theoretical and experimental systematical errors and it is therefore essential to improve on these errors.

In case LHC observes a significant excess in the mode $t\bar{t}Z$, our interpretation predicts several possible complementary signals, in particular a drastic change in the ratio tR/tL, which would allow to confirm unambiguously the origin of this excess.

ILC would provide an ideal tool pushing the accuracy by one order of magnitude and would allow to study very precisely and disentangle the various photon and Z top couplings. It would therefore allow to fully reveal the underlying mechanism by separating the vector, tensor and axial couplings of this particle [11].

**Acknowledgements**

I warmly thank Abdelhak Djouadi for encouraging this work and providing very useful suggestions.